
%
\magnification=\magstep1
\vsize=23truecm
\hsize=15.5truecm
\hoffset=.2truecm
\voffset=.8truecm
\parskip=.2truecm

\font\ti=cmbx10 scaled\magstep1
\font\eightrm=cmr8
\font\ninerm=cmr9
\def \es{e^{-\kappa \sigma}}
\def \ra{\rightarrow}

\def\br{\hfill\break\noindent}

\def \lra{\leftrightarrow}
\def \ot {\otimes}
\def \g5{\gamma_5}
\def \ra{\rightarrow}
\def \l4{\Bigl( {\rm Tr}( KK^*)^2-({\rm Tr}KK^*)^2\Bigr)}
\def \bp{\oplus }
\def \k2{{\rm Tr}KK^*}
\def \slash#1{/\kern -6pt#1}
\def \di{\slash{\partial}}
\def \Di{\slash{\nabla}}

\def\u{\underline}

\def \G11{\Gamma_{11}}
\def\m {\cal M}
\def\n {\cal N}

\def \s {SO(10)}
\def \a {\alpha }
\def \b {\beta }
\def \ah {\hat{\alpha }}
\def \bh {\hat {\beta }}
\def \ot {\otimes}
\def \g5{\gamma_5}
\def \ra{\rightarrow}
\def \l4{\Bigl( {\rm Tr}( KK^*)^2-({\rm Tr}KK^*)^2\Bigr)}
\def \bp{\oplus }
\def \k2{{\rm Tr}KK^*}
\def \slash#1{/\kern -6pt#1}
\def \di{\slash{\partial}}

\pageno=1
%
\baselineskip=12truept
\footline={\hfill}
\centerline{\ti Particle Physics Models, Grand Unification }
\centerline{\ti and Gravity in Non-Commutative Geometry}
\vskip1.2truecm
\centerline{  A. H. Chamseddine$^{1}$ \footnote*
{\ninerm Supported by the Swiss National Foundation (SNF)}
and J. Fr\"ohlich$^{2}$ \footnote{\dag}
{\ninerm Permanent address: Theoretische Physik, ETH, CH 8093
Z\"urich Switzerland }}
\vskip.8truecm
\centerline{$^{1}$ Theoretische Physik, Universit\"at Z\"urich, CH
8001 Z\"urich Switzerland}
\centerline{$^{2}$ Institut des Hautes Etudes Scientifique,
F-91440 Bures-Sur-Yvette, France }
\vskip1.2truecm
\centerline{\sl Dedicated to Abdus Salam on his 65th birthday}
\vskip.5truecm
\centerline{\bf Abstract}
\vskip.3truecm
\noindent
We review the construction of particle physics models in the
framework of non-commutative geometry. We first give simple
examples, and then progress to outline the Connes-Lott
construction of the standard
Weinberg-Salam model and our construction of the  SO(10) model.
We then discuss the analogue
of the Einstein-Hilbert action and gravitational matter couplings.
Finally we speculate on some experimental signatures of predictions
specific to the non-commutative approach.
\baselineskip=12truept
\footline={\hss\eightrm\folio\hss}
\vskip.7truecm
{\bf\noindent 1. Introduction}
\vskip.15truecm

\noindent
The Weinberg-Salam model [1] of electroweak interactions is a
milestone in the search for unity of all fundamental interactions.
But although this model has passed all  experimental tests
at present energies, many challenges  remain. To name just a
few, we have to understand:\br
a-The role of the Higgs field necessary in the spontaneous
breakdown of the $SU(2)\times U(1)$ gauge symmetry.\br
b-The fermionic mass matrices and family mixing, the gauge
coupling constants, the mass and vacuum expectation value
(vev) of the Higgs field.\br
c-Unifying gravity with the strong and electroweak interactions
in a renormalizable theory.\br
There are many attempts to solve these problems using
schemes such as grand unification, Kaluza-Klein compactification
and string theory, all with and without supersymmetry. The virtues
and shortcomings of these lines of research are now well known.

During the past few years, Connes has proposed a construction
of particle physics models based on
his formulation of non-commutative geometry [2]. This method addresses
point a- raised above, in that it predicts the existence of the
Higgs field and gives it a geometrical meaning [3]. This article is
a short review of  Connes'
non-commutative construction and intended for particle physicists.
The mathematics used here will be
the  minimum needed. For the more mathematically oriented reader
we refer to some of the available reviews [4]. Our
plan is as follows. In section 2 we introduce the non-commutative
construction and give simple examples. In section 3 we review the
derivation of the standard model and in section 4 the grand unified
SO(10) model.
In section 5 we describe an  analogue of the Einstein-Hilbert action
and the gravitational matter couplings, and, under  a natural
geometrical assumption, obtain some predictions for the top quark
mass and the Higgs mass.


{\bf \noindent 2. The non-commutative construction }
\vskip.15truecm
\noindent
Connes' non-commutative geometry is very general [2].
A non-commutative geometry is specified
by the triple $({\cal A}, h,D)$, where $h$ is a Hilbert space,
$\cal A$ is an involutive algebra of operators on $h$, and $D$
is an unbounded self-adjoint operator on $h$. Let $\Omega^.$ be
the $Z$ graded differential algebra of universal forms over $R$ or $C$:
$\Omega^.=\oplus_n \Omega^n $, where ${\cal A}=\Omega^0$ and
$\Omega^n $ is the space of n-forms with operations i) $d: \Omega^n
\ra \Omega^{n+1}$, ii)$m: \Omega^n \ot \Omega^m \ra \Omega^{n+m}
$. The algebra of universal forms over $\cal A$, $\Omega^.(\cal
A)$, is generated by $f$ and $df$, where $f\in \cal A$.
The operator $d$ obeys Leibnitz rule, $d(fg)=(df)g+f(dg)$, where
$f,g\in \cal A$, and $d^2=0$. An n-form in $\Omega^n(\cal A)$
is given by $\sum_i a_0^ida_1^i\cdots da_n^i; \qquad a_0^i,\cdots
a_n^i \in \cal A$. An involutive representation of $\Omega^.(
\cal A)$ on $h$ is provided by the map $\pi :\Omega^.({\cal A})\ra
B(h)$ defined by
$$
\pi (a_0da_1\cdots da_n )=a_0[D,a_1]\cdots [D,a_n], \eqno(2.1)
$$
where $B(h)$ is the algebra of bounded operators on $h$. The
non-commutativity resides in the fact that $ab$ is not
necessarily equal, up to a sign, to $ba$. Let $E$ be a
vector bundle determined by the vector space $\cal E$ of
its sections. We will be mainly interested in the case
${\cal E}=\cal A$.  Let $\rho $ be a self-adjoint element in
the space $\Omega^1(\cal A)$. It determines a connectiion with
curvature $\theta
=d\rho +\rho^2 \ \in \Omega^2 (\cal A)$. The Yang-Mills
action functional is obtained using the Dixmier trace which
permits the definition of integration and volume elements
in non-commutative geometry. We set (see [2,3])
$$
I_{\rm YM}={\rm Tr}_w \bigl( \theta^2 D^{-4}\bigr), \eqno(2.2)
$$
The same quantity can be defined using the heat kernel expansion
(see [5]);i.e.,
$$
{\rm lim}_{\epsilon \ra 0}{{\rm Tr}_H(\theta^2 e^{-\epsilon D^2})
\over {\rm Tr}_H (e^{-\epsilon D^2})}. \eqno(2.3)
$$
We illustrate these notions with two simple examples.

\noindent
1-Let ${\cal A}_1=C^{\infty}(M)$, the algebra of functions on a
four-dimensional Riemannian manifold $M$, $h$ the Hilbert
space of spinors $L^2 (M,\sqrt g d^4x)$ and $D_1=\di $, the
Dirac operator on $h$. The one-form $\rho =\sum_i a^idb^i$
has the image under $\pi $
$$\pi (\rho )=\sum_i a^i[D,b^i]=\sum_i a^i\di b^i\equiv
\gamma^{\mu}A_{\mu}. \eqno(2.4).
$$
Similarly for the two-form $d\rho $ we have
$$
\pi (d\rho )=\sum_i [D,a^i][D,b^i]=\sum_i \di a^i\di b^i .\eqno(2.5)
$$
The curvature $\pi (\theta )=\pi (d\rho )+\pi (\rho )^2$
is then given by
$$
\pi (\theta )={1\over 2}\gamma^{\mu\nu}F_{\mu\nu}\ +X ,\eqno(2.6)
$$
where $\gamma^{\mu\nu}={1\over 2}(\gamma^{\mu}\gamma^{\nu}
-\gamma^{\nu}\gamma^{\mu})$,
$X=g^{\mu\nu}(A_{\mu}A_{\nu}
+\sum_i \partial_{\mu}a^i\partial_{\nu}b^i)$ is an "auxiliary field"
and $F_{\mu \nu}$
is the field strength of $A_{\mu}$. Notice that
$\pi (d\rho )_{\rho =0}=g^{\mu\nu}\sum_i \partial_{\mu}a^i
\partial_{\nu}b^i \ne 0$, is a scalar function. This is
the reason behind the presence of the auxiliary field in
$\pi (\theta )$. It is possible to work instead with the
space ${\Omega^2({\cal A})\over {\rm Ker}\pi +d{\rm Ker}\pi}$,
but we will not do this now. The Yang-Mills action becomes
$$
I_{YM}=\int d^4 x (-{1\over 4}F_{\mu\nu}F^{\mu\nu} \ + X^2).\eqno(2.7)
$$
After eliminating the auxiliary field $X$ by its equation
of motion, it decouples from the action.

\noindent
2-For a two point space, we take ${\cal A}_2=C\oplus C$, and
$h=C^N\oplus C^N$ and the Dirac operator is $D_2=\pmatrix
{0&K\cr K^*&0\cr}$, where $K$ is an $N\times N$ matrix.
The elements $a\in {\cal A}_2$ have the representation
$a\ra {\rm diag}(a_1,a_2)$, $a_1, a_2\in C$. Then
$$
\pi (\rho )=\sum_i a^i[D,b^i]=\pmatrix{0&K\phi \cr K^*\phi^*&0\cr},\eqno(2.8)
$$
where $\phi =\sum_i a_1^i(b_2^i-b_1^i)$ and $\phi^* =\sum_i
a_2^i (b_1^i-b_2^i)$. Then $\pi (d\rho )=\sum_i [D,a^i][D,b^i]$
is equal to
$$
\pi (d\rho )=-\pmatrix {KK^* (\phi +\phi^*)&0\cr 0&K^*K(\phi +\phi^*)\cr}.
\eqno(2.9)
$$
The Yang-Mills action is easily calculated to be
$$
{\rm tr} (\theta^2 )=2{\rm Tr}(KK^*)^2 \bigl( \vert \phi -1\vert^2
-1\bigr)^2 ,\eqno(2.10)
$$
It is seen to be of the same form as the Higgs potential
for a scalar field $\phi $ and is positive definite.
Notice that $[D,a]=\pmatrix{0&K(b_2-b_1)\cr K^*(b_1-b_2)&0}$
is a difference operator in the discrete space.

{\bf\noindent 3. The standard Weinberg-Salam model}
\vskip.15truecm
With the simple tools introduced in the last section, we now
show that it is possible to construct realistic action functionals.
Not all models are possible, but for those ones which are, the Higgs
structure is fixed. For  lack of space we shall only
describe the standard Weinberg-Salam model in this section and
the grand unification
SO(10) model in the next section. Our method is a modified variant of
Connes' construction (simplifying some computations
[5]).

Combining examples 1 and 2, let the algebra be ${\cal A}={\cal A}_1
\ot {\cal A}_2$ acting on the Hilbert space $h=h_1\ot h_2$, where
${\cal A}_1=C^{\infty}(M)$, considered before, and ${\cal A}_2=
M_2(C)\oplus M_1(C)$ the algebras of $2\times 2$ and $1\times 1$
matrices. The Hilbert space is that of spinors of the form
$L=\pmatrix{l\cr e\cr}$ where $l$ is a doublet and $e$ is a
singlet. The spinor $L$ satisfies the chirality condition
$\g5 \ot \Gamma_1 L=L$, where $\Gamma_1 ={\rm diag}(1_2,-1)$
is the grading operator. This implies that $l=l_L$ is left-handed
and $e=e_R$ is right-handed, and so we can write $l_L=\pmatrix
{\nu_L\cr e_L\cr}.$ The Dirac operator is $D=D_1\ot 1 +\Gamma_1
\ot D_2$, so that
$$
D_l=\pmatrix{\di \ot 1_2& \g5 M_{12}\ot k\cr
\g5 M_{21}\ot k^* &\di \cr}, \eqno(3.1)
$$
where $M_{21}=M_{12}^*$ and $k$ is a family mixing matrix.
The geometry is that of a four-dimensional manifold $M$
times a discrete space of two points. The column $M_{12}$
in $D$, the vev of the Higgs
field,  is
taken to be $M_{12}=\mu\pmatrix{0\cr 1}\equiv H_0$. The elements
$a\in \cal A$ have the representation $a\ra
{\rm diag}(a_1,a_2)$ where $a_1$ and $a_2$ are $2\times 2$
and $1\times 1$ unitary matrix-valued functions, respectively.
The self-adjoint one-form $\rho $ has the representation
$$
\pi_l (\rho )=\pmatrix{A_1\ot 1_3 &\g5 H\ot k\cr
\g5 H^*\ot k^* &A_2\ot 1_3\cr}, \eqno(3.2)
$$
where $A_1=\sum_i a_1^i\di b_1^i $, $A_2=\sum_ia_2^i\di b_2^i $
and $H=H_0+\sum_i a_1^iH_0b_2^i$. In a world without quarks,
the generalized tracelessness condition ${\rm Tr}(\Gamma_1 \pi
(\rho ))=0$ allows the gauge fields to be written in the form
$A_1=-{i\over 2 }g_2 \sigma^a A_a +ig_1B$, $A_2=2ig_1B$ where
$g_1,g_2$ are the U(1) and SU(2) gauge couplings. The leptonic
action $<L, (D+\rho )L>$ gives the correct lepton couplings to
the gauge and Higgs fields. However, to be realistic, the quarks
and the SU(3) gauge group must be introduced.
This can be achieved by taking a bimodule structure relating
two algebras $\cal A$ and $\cal B$, where the algebra $\cal B$
is taken to be $M_1 (C)\oplus M_3(C)$, commuting with the action
of $\cal A$, and the mass matrices in the Dirac operator are taken
to be zero when acting on elements of $\cal B$. Then the one-form
$\eta $ in $\Omega^1 (\cal B)$ has the simple form $\pi_l(\eta )
=B_1{\rm diag}(1_2,1)$, where $B_1$ is a $U(1)$ gauge field associated
with $M_1(C)$.
The quark Hilbert space is that of the
spinor $Q=\pmatrix{u_L\cr d_L\cr d_R \cr u_R}$. The representation
of $a\in \cal A$ is: $a\ra {\rm diag}(a_1,a_2,\overline {a_2})$
where $a_1$ is a $2\times 2$ matrix-valued function and $a_2$
is a complex-valued function. The Dirac operator acting on the
quark Hilbert space is
$$
D_q=\pmatrix{\gamma^{\mu} (\partial_{\mu} +\ldots )\ot 1_2
\ot 1_3 &\g5  \ot M_{12} \ot k{'}&\g5 \ot \tilde{ M_{12}}\ot k{''}\cr
\g5  \ot M_{12}^* \ot k^{'*}&\gamma^{\mu}(\partial_{\mu} +\ldots )
\ot 1_3 &0\cr
\g5 \ot \tilde {M_{12}}^*\ot k^{''*}&0&\gamma^{\mu}
(\partial_{\mu} +\ldots )\ot 1_3\cr}\ot 1_3, \eqno(3.3)
$$
where $k'$ and $k^{''}$ are $3\times 3$ family mixing matrices,
and $\tilde {M_{12}}=\mu \pmatrix{1\cr 0\cr}$.
Then the one-form in $\Omega^1(\cal A)$ has the representation
$$
\pi_q(\rho )=\pmatrix{A_1\ot 1_3& \g5 H\ot k'&\g5 \tilde H\ot
k^{''}\cr \g5 H^*\ot k^{'*}&A_2\ot 1_3&0\cr
\g5 {\tilde H}^*\ot k^{''*}&0&\overline {A_2}\ot 1_3\cr},\eqno(3.4)
$$
where $\tilde H_a=\epsilon_{ab}H^b$.
On the algebra $\cal B$ the Dirac operator has zero mass matrices,
and the one form $\eta $ in $\Omega^1(\cal B) $ has the
representation $\pi_q(\eta )= B_2{\rm diag}(1_2,1,1)$ where
$B_2$ is the gauge field associated with $M_3(C)$.
Imposing the unimodularity condition on the algebras $\cal A$
and $\cal B$ relates the U(1) factors in both algebras [3]:
${\rm tr}(A_1)=0$, $A_2=B_1=-{\rm tr}B_2={i\over 2}g_1B$.
We can then write
$$\eqalign{
A_1&=-{i\over 2}g_2 A^a\sigma_a \cr
B_2&=-{i\over 6}g_1B -{i\over 2}g_3 V^i\lambda_i \cr}
$$
where $g_3$ is the SU(3) guge coupling
constant, and $\sigma^a $ and $\lambda^i $ are the Pauli
and Gell-Mann matrices, respectively.
The fermionic action for the leptons is
$$
<L, (D+\rho +\eta )L>=\int d^4x \sqrt g \Bigl(
\overline L \bigl( D_l +\pi_l (\rho )+\pi_l (\eta )\bigr)L
\Bigr), \eqno(3.5)
$$
and, for the quarks it is
$$
<Q, (D+\rho +\eta )Q>=\int d^4x \sqrt g \Bigl(
\overline Q \bigl( D_q +\pi_q (\rho )+\pi_q (\eta )\bigr)Q
\Bigr),  \eqno(3.6)
$$
and these can be easily checked to reproduce the standard
model lepton and quark interactions with the correct hypercharge
assignments.

The bosonic actions are the square of the curvature in the
lepton and  quark spaces, and are given, respectively, by
$$ \eqalign{
I_l&={\rm Tr}(C_l(\theta_{\rho}+\theta_{\eta})^2 D_l^{-4}) \cr
I_q&={\rm Tr}(C_q(\theta_{\rho}+\theta_{\eta})^2 D_q^{-4})\cr}.\eqno(3.7)
$$
To compute the bosonic action, we use a general formula,
derived in [5], based on a Dirac operator where the discrete space
has N points:
$$
D =\pmatrix{\di \ot 1\ot 1 &\g5 \ot M_{12}\ot K_{12} &
\ldots &\g5 \ot M_{1N} \ot K_{1N}\cr
\g5 \ot M_{21}\ot K_{21} &\di \ot 1\ot 1&\ldots &\g5 \ot M_{2N}
\cr \vdots &\vdots &\ddots &\vdots \cr
\g5 \ot M_{N1} \ot K_{N1}& \g5\ot M_{N2}\ot K_{N2} &\ldots &
\di \ot 1 \cr},\eqno(3.8)
$$
where the $K_{mn}$ are $3\times 3$ matrices commuting with the $a_i$
and $b_i$. The Yang-Mills action associated with this operator is
$$\eqalign{
I_b&=\sum_{m=1}^N Tr\Bigl( {1\over 2}  F_{\mu\nu}^m F^{\mu\nu m}
-\Bigl\vert  \sum_{p\not= m}\vert K_{mp}\vert^2
\vert \phi_{mp}+M_{mp}\vert^2
-(Y_m +X'_{mm})\Bigr\vert^2 \cr
& + \sum_{p\not=m} \vert K_{mp}\vert^2
\Bigl\vert \partial_{\mu}
(\phi_{mp}+M_{mp})+A_{\mu m}(\phi_{mp}+M_{mp})-(\phi_{mp}+M_{mp})
A_{\mu p}\Bigr\vert^2 \cr
& - \sum_{n\not=m}\sum_{p\not=m,n}\Bigl\vert
K_{mp}K_{pn}\bigl(
(\phi_{mp}+M_{mp})(\phi_{pn}+M_{pn})-M_{mp}M_{pn}\bigr) -X_{mn}\Bigr)
\Bigl\vert^2
\Bigr), \cr}\eqno(3.9)
$$
where the $A^m$ are the gauge fields in the $m-m$ entry of $\pi
(\rho )$ and $\phi_{mn} $ are the scalar fields in the $m-n$ entry
of $\pi (\rho )$. The $X_{mn}$, $X'_{mn}$ and $Y_m$ are fields whose
unconstrained elements are auxiliary fields that can be
eliminated from the action. Their expressions in terms of the
$a^i$ and $b^i $ are
$$\eqalignno{
X_{mn}&=\sum_i a_m^i \sum_{p\ne m,n} K_{mp}K_{pn} (M_{mp}M_{pn}
b_n^i -b_m^i M_{mp}M_{pn} ), \qquad m\ne n, &(3.10)\cr
X'_{mm}&=\sum_i a_m^i\di^2b_m^i +(\partial^{\mu}
A_{\mu}^m+A^{\mu m}A_{\mu}^m ),&(3.11) \cr
Y_m&=\sum_{p\ne m}\sum_i a_m^i\vert K_{mp}\vert^2 \vert M_{mp}
\vert^2 b_m^i. &(3.12)\cr}
$$
Using Eqs (3.9)-(3.12) for the leptons and quarks seperately,
yields an action containing
the kinetic terms for the U(1), SU(2) and SU(3) gauge fields,
as well as the kinetic energy and potential of the Higgs field.
The most complicated step is  the elimination of the
auxiliary fields, but this only changes the coefficients
of the Higgs potential, not its form.
By writing $C_l ={\rm diag}(c_1,c_1,c_2)$ and $C_q={\rm diag}
(c_3,c_3,c_4,c_4)$, the bosonic action depends on the
constants $c_1,c_2,c_3,c_4,g_1,g_2,g_3$ as well as on the Yukawa
couplings. Normalizing the kinetic energies of the
$SU(3)$, $SU(2)$ and $U(1)$ gauge fields fixes three of the constants
$c_1,\ldots ,c_4$ in terms of $g_1,g_2,g_3$. In the special case
when $c_1=c_2=c_3=c_4$, one gets a constraint on the gauge
coupling constants as well as fixed values for the Higgs mass
and top quark mass [3]. These relations cannot be maintained
after quantization, as can be seen from the renormalization
group equations for the coupling constants and the masses [6].
We shall not assume any such relations among the $c's$.
The Higgs sector
is then parametrized in terms of two parameters $\lambda $ and
$m$ which are functions of
of the $c's$, $k,s$ and $H_0$.
The bosonic part of the standard model becomes
$$\eqalign{
L_b &= -{1\over 4} \Bigl ( F_{\mu\nu}^3 F^{\mu\nu 3}
+F_{\mu\nu}^2 F^{\mu\nu 2} +F_{\mu\nu}^1 F^{\mu\nu 1} \Bigl) \cr
&\qquad +D_{\mu} (H+M_{12})^* D_{\nu}(H+M_{12}) g^{\mu\nu} \cr
&\qquad -{\lambda \over 24} \Bigl\vert \vert H+M_{12}\vert^2
-\vert M_{12}\vert^2 \Bigr\vert^2  \cr}.\eqno(3.13)
$$
The cosmological constant comes out to be zero, naturally, at the
classical level.

{\bf \noindent 4. SO(10) unification model.}
\vskip.3truecm
\noindent
The way the strong interactions are introduced in the
standard model suggests that a more unified picture would
be preferable. The starting point is the Hilbert space
of spinors and the Dirac operator acting on this space.
The arrangement of fermions determines the structure of the
discrete space. We place the fermions in the $16_s$ spinor
representation of $SO(10)$ [7]. This is a 32-component spinor
subject to the space-time and SO(10) chirality
$$
\eqalign{
(\g5 )_{\a}^{\b}\psi_{\b\ah}&=\psi_{\a\ah} \cr
(\G11 )_{\ah}^{\bh}\psi_{\a\bh}&=\psi_{\a\ah}.\cr}\eqno(4.1)
$$
where $\G11 =-i\Gamma_0\Gamma_1 \cdots \Gamma_9 $.
This reduces the independent spinor components to two for
the space-time indices, and to sixteen for the $\s $ indices.
The general fermionic action is given by
$$
\overline {\psi_{\a\ah}^p}\bigl( \di +A^{IJ}\Gamma_{IJ}\bigr)_{\a\ah}
^{\b\bh}\psi_{\b\bh}^p \ +\psi_{\a\ah}^{Tp}C^{\a\b}H_{\ah\bh}^{pq}
\psi_{\b\bh}^q, \eqno(4.2)
$$
where $C$ is the charge conjugation matrix,  $p, q=1,2,3$ are
family indices, and $H$ is some appropriate combination of
Higgs fields breaking the subgroup $SU(2)\times U(1)$
of $SO(10)$ at low energies. An exception of a Higgs
field that  breaks the symmetry at high energies and yet couples to
fermions is the one
that gives a Majorana mass to the right handed neutrinos . The
other Higgs fields needed to break the $SO(10)$ symmetry at
high energies should not couple to the fermions so as not to
give the quarks and leptons super heavy masses.
The simplest picture corresponds to the
spinor $\Psi =\pmatrix{P_+\psi \cr P_+\psi \cr P_-\psi^c\cr}$ where
$\psi^c\equiv BC\overline{\psi^T}$,  $B$ is the SO(10) conjugation
matrix satisfying $B^{-1}\Gamma_IB=-\Gamma_I^T$ and $P_{\pm}
={1\over 2}(1\pm \Gamma_{11})$. However, it
turns out that the model associated with this arrangement,
although elegant, is not realistic, because
the Cabbibo angle vanishes [8]. The correct model is the one with the
spinor
$$
\Psi =\pmatrix{P_+\psi \cr P_+\psi \cr
P_-\psi^c \cr P_-\psi^c \cr \lambda \cr \lambda^c \cr},\eqno(4.3)
$$
where $\lambda $ is a singlet fermion that will couple
to the right-handed neutrino in the $16_s$.
The algebra $\cal A$ is equal to ${\cal A}_1 \ot {\cal A}_2$ where
${\cal A}_1 =C^{\infty }(M)$, and
$$
{\cal A}_2\equiv P_+{\rm Cliff}\bigl( SO(10) \bigr)P_+\bp R. \eqno(4.4)
$$
The involutive map $\pi $ is taken to be
$$
\pi (a)=\pi_0(a)\bp \pi_0(a)\bp \overline\pi_0 (a) \bp\overline\pi_0 (a)
\bp\overline\pi_0 (a)\bp \pi_1(a)\bp \pi_1(a), \eqno(4.5)
$$
acting on the Hilbert space $\tilde h=h_1\ot (h_2^{(1)}\bp
\cdots \bp h_2^{(6)})$ where $h_2^{(i)}\cong h_2, \quad
i=1,\cdots 4$,  $h_2$ is the 32 dimensional Hilbert space
on which ${\cal A}_2$ acts, and $h_2^{(i)}\cong C,\quad i=5,6$.
Let $h$ be the subspace of $\tilde h$ which is the image of
the orthogonal projection onto elements of the form (4.3). On
$\tilde h$ the self-adjoint Dirac operator has the form
(3.8), for N=6. From Eq (4.5) we have the permutation symmetry $1\lra 2$,
$3\lra 4$, $5\lra 6$, and the conjugation symmetry $1\lra 3$,
and the one-form $\pi (\rho )$ reads
$$
\pi (\rho )=\pmatrix{A &\g5 {\m}  K_{12}&
\g5 {\n} K_{13}&\g5 {\n} K_{14}&\g5 H K_{15}&
\g5 H K_{16}\cr \g5 {\m} K_{12} &A &
\g5 {\n} K_{23} &\g5 {\n} K_{24}&\g5  H K_{25}
&\g5  H  K_{26}\cr
\g5 {\n}^*  K_{31}&\g5 {\n}^*  K_{32} &B \overline A B^{-1}
 &\g5 {\m }{'} K_{34} &
\g5 H' K_{35}&\g5 H'  K_{36}\cr
\g5 {\n}^*  K_{41}&\g5 {\n}^* K_{42}&\g5
{\m }^{'} K_{43}&B \overline A B^{-1}  &\g5 H'  K_{45}&\g5  H' K_{46}\cr
\g5 H^* K_{51}&\g5 H^* K_{52}&\g5 H^{'*} K_{53}
&\g5 H^{'*} K_{54}&0&0\cr
\g5 H^*  K_{61}&\g5 H^* K_{62}&\g5 H^{'*} K_{63}&
\g5 H^{'*} K_{64}&0& 0\cr},\eqno(4.6)
$$
where the new functions $A$, $\m $, $\n $ and $H$ are given in terms
of the $a^i$ and $b^i$ by
$$\eqalign{
A&=P_+(\sum_i a^i \di b^i )P_+ \cr
{\m}+{\m}_0 &=P_+(\sum_i a^i {\m}_0 b^i )P_+ \cr
{\n}+{\n}_0 &=P_+(\sum_i a^i {\n}_0 B\overline {b^i}B^{-1})P_- \cr
H+H_0 &=P_+(\sum_i a^i H_0 b^{'i})\cr}\eqno(4.7)
$$
and ${\m}^{'}=B\overline {\m} B^{-1} $, $H'=B\overline H $.
We can expand these
fields in terms of the $SO(10)$ Clifford algebra as
follows:
$$\eqalign{
A &=P_+ \bigl( ia +a^{IJ}\Gamma_{IJ} +i a^{IJKL}\Gamma_{IJKL}
\bigr) P_+ \cr
{\m} &= P_+\bigl( m+i m^{IJ}\Gamma_{IJ} +m^{IJKL}\Gamma_{IJKL}
\bigr)P_+ \cr
{\n} &=P_+\bigl( n^I \Gamma_I +n^{IJK}\Gamma_{IJK}
+n^{IJKLM}\Gamma_{IJKLM}\bigr) P_-. \cr}
\eqno(4.8)
$$
The self-adjointness condition on
$\pi (\rho ) $ implies, after using the hermiticity of the
$\Gamma_I $ matrices, that all the fields $a$ and $m$
 appearing in the expansion
of $A, \m $  are real, because both are self-adjoint,
while those in $\n $ are complex. Imposing the reality condition
on the coefficients of the Clifford algebra expansion of the gauge
field $A$ forces $a=0=a^{IJKL}$, reducing the gauge group from
U(8) to SO(10).
The symmetry breaking pattern that breaks the gauge group $\s $ must be
coded into the Dirac operator $D$. The Higgs fields at
our disposal are $\m $,  $\n $ and $H$. In terms of $\s $ representations
these are $\u 1,\ \u {45},\ \u {210}$ in $\m $,  complex $
\u {10}, \ \u {120} $ and $\u {126} $ in $\n $ and $\u {16}_s$
in $H$.
To be explicit we shall work in a specific $\Gamma
$ matrix representation.
The $32\times 32 \ \Gamma $ matrices are represented in terms of tensor
products of five sets of Pauli matrices $\sigma_i ,\tau_i ,\eta_i
,\rho_i ,\kappa_i $ where $i=1,2,3$.
The $\Gamma $ matrices are given by
$$\eqalign{
\Gamma_i &=\kappa_1 \rho_3\eta_i ,\qquad
\Gamma_{i+3}= \kappa_1 \rho_1 \sigma_i \cr
\Gamma_{i+6}&= \kappa_1 \rho_2\tau_i ,\qquad
\Gamma_0 =\kappa_2 ,\qquad
\Gamma_{11}=\kappa_3 \cr}\eqno(4.9)
$$
where $i=1,2,3,$
and where we have omitted
the tensor product symbols. In this basis, an $\s $ chiral spinor
will take the form $\psi_+ =\pmatrix{\chi_+ \cr 0\cr }$
where $\chi $ is a $\u {16}_s $.
The $\s $ conjugation matrix is defined
by $ B\equiv -\Gamma_1\Gamma_3\Gamma_4\Gamma_6\Gamma_8 $ which, in
the basis of equation (4.9), becomes
$$
B=\kappa_1 \rho_2\eta_2\tau_2\sigma_2 \equiv \kappa_1 b \eqno(4.10)
$$
where the matrix $b=\rho_2\eta_2\tau_2\sigma_2 $ is the conjugation
matrix in the space of the
sixteen component spinors. The action
of $B$ on a chiral spinor is then $B\psi_+
=\pmatrix {0 \cr b\chi_+ \cr}$.
The advantage of this system of matrices is that
$bC \overline {\chi_+}^T $, have the same form as $\chi_+$
but is right-handed not left-handed.
To correctly
associate the components of $\chi_+ $ with  quarks and leptons,
we consider the action of the charge operator [7] on $\chi_+ $:
$$\eqalign{
Q&=-{1\over 6}(\sigma_3 +\tau_3 +\rho_3\tau_3\sigma_3) +{1\over 2}\eta_3
\cr} \eqno(4.11)
$$
which gives
$$
Q\chi_+ ={\rm diag }(0,{2\over 3},{2\over 3},{2\over
3},-1,-{1\over 3}, -{1\over 3},-{1\over 3}, {1\over 3},{1\over 3},
{1\over 3},1,-{2\over 3},-{2\over 3},-{2\over 3},0)\chi_+ \eqno(4.12)
$$
Thus the components of the left handed spinor $\chi_+ $ are  written as
the column
$$\eqalign{
\chi_+&=( n_L, u_L^1, u_L^2,
u_L^3, e_L, d_L^1, d_L^2, d_L^3,\cr
& \qquad \qquad  -(d_R^3)^c,
(d_R^2)^c, (d_R^1)^c, -(e_R)^c, (u_R^3)^c,
-(u_R^2)^c, -(u_R^1)^c, (n_R)^c)}\eqno(4.12)
$$
where the $c$ in this equation stands for the usual charge
conjugation, eg. $d^c =C \overline d^T $. The upper and lower
components in $\chi $ are mirrors, with the signs chosen so
that the spinor $bC\overline {\chi_+}^T $ has exactly the same
form as $\chi_+ $, but with the left-handed and right handed signs,
$L $ and $R$, interchanged.
We now specify the vevs  ${\m}_0 $,
${\n}_0 $ and $H_0$. The group $\s $ is broken at high energies by
$\m $ which contains the representations $\u {45} $ and
$\u {210} $. By taking the vev of the $\u {210}$ to be
${\m}^{0123}=O(M_G) $, the $\s $ symmetry is broken to $SO(4)\times
SO(6) $ which is isomorphic to $SU(4)_c\times
SU(2)_L\times SU(2)_R $. The $SU(4)_c$ is further broken to
$SU(3)_c\times U(1)_c $ by the vev of the $\u {45}$. Therefore we
write [8]
$$\eqalign{
P_+{\m}_0 P_+ &=P_+\Bigl( M_G \Gamma_{0123} -iM_1 (\Gamma_{45}
+\Gamma_{78}+\Gamma_{69}) \Bigr)P_+ \cr
&={1\over 2}(1+\kappa_3)\Bigl(- M_G \rho_3 +M_1 (\sigma_3 +\tau_3+
\rho_3\tau_3\sigma_3 ) \Bigr) \cr}.\eqno(4.13)
$$
Therefore ${\m}_0 $ breaks $\s $ to $SU(3)_c\times U(1)_c
\times SU(2)_L\times SU(2)_R $ which is also of rank five. The rank
is reduced by giving a vev to the components of
$\u {126} $ that couple to the right-handed neutrino.Therefore
the vev of ${\n}_0$ must contain the term
$$
M_2({1\over 2^5}) (\kappa_1 +i\kappa_2 )(\rho_1 +i\rho_2 )(\eta_1 +i\eta_2 )
(\tau_1 +i\tau_2 )(\sigma_1 +i\sigma_2 ) \eqno(4.14)
$$
The vev of ${\n}_0$ break $U(1)_c\times SU(2)_R$ to $U(1)_Y$, and the
surviving group would
be the familiar $SU(3)_c\times SU(2)_L\times U(1)_Y$. This
breaking is also obtained for an $H_0$ whose vev is
$H_0=M_3\pmatrix{0\cr \vdots \cr 0\cr 1\cr}$. As we shall explain
shortly, $M_1, M_2$ and $M_3$ must be related for the model to
be consistent.
The only generators that leave $ {\m}_0 $,
the part of ${\n}_0 $ given by 4.14 and $H_0$ invariant
are those of the standard model.
The eight $SU(3)$ generators
are given by $(1-\rho_3\tau_3)\sigma_i $,
$(1-\rho_3\sigma_3)\tau_i $, $\rho_3 (\tau_1\sigma_1 +\tau_2\sigma_2)$
and $\rho_3 (\tau_2\sigma_1 -\tau_1\sigma_2) $. The $SU(2)_L$
generators are ${1\over 2}(1\pm \kappa_3\rho_3)\eta^i$.
Finally the $U(1)_Y$
generator is
related to the charge operator $Q$ by
$Q={1\over 2} Y+T_L^3  $,
where the action of the $SU(2)_L$ isospin $T_L^3$ on $\chi_+ $ is given by
$T_L^3={1\over 2}(1+\rho_3)\eta_3 $.

For the last stage of symmetry breaking of $SU(2)_L\times
U(1)_Y$ we can use the field $\n $ which contains the
compex representations $\u {10} $, $\u {120} $ and $\u {126} $. The
most general vev that preserves the group $SU(3)_c \times U(1)_Q$ is
$$\eqalign{
P_+{\n}_0 P_-\kappa_1 &={1\over 2}(1+\kappa_3 )\Bigl( s+p\rho_3\eta_3
+a\rho_3 +a'\eta_3 \cr
&\qquad +(b'+b\rho_3\eta_3 +e\eta_3 +f\rho_3 )(\sigma_3 +\tau_3
+\rho_3\tau_3\sigma_3 )\cr
&\qquad +M_2 ({1\over 2^5})(\rho_1 +i\rho_2 )(\eta_1 +i\eta_2 )(\tau_1
+i\tau_2 )(\sigma_1 +i\sigma_2 )
\Bigr), \cr}\eqno(4.15)
$$
where all terms containing $\eta_3 $ break $SU(2)_L \times
U(1)_Y $ and $s, p, a, a', b, b', e, f$ are $O(M_W)$.
The fermionic action is simply given by
$$\eqalign{
I_{{\rm f-mass}}&=<\Psi ,(D+\rho )\Psi> \cr
&=-\int d^4 x \Bigl( \bigl( (s+p +3(e+f))K_{(pq)}
+(a+a'+3(b+b'))K_{[pq]} \bigr) \overline {N_R^p} N_L^q \cr
&\qquad +\bigl( (s+p-(e+f))K_{(pq)} +(a+a'-(b+b'))K_{[pq]}
\bigr) \overline {u_R^p} u_L^q \cr
&\qquad +\bigl( (s-p-3(e-f))K_{(pq)} +(a-a'-3(b-b'))K_{[pq]}
\bigr) \overline {e_R^p} e_L^q \cr
&\qquad +\bigl( (s-p +e-f)K_{(pq)} +(a-a'+b-b')K_{[pq]}\bigr)
\overline {d_R^p} d_L^q \cr
&\qquad +\bigl( \sqrt 2 M_3 K_{pq}^{'}\overline {N_R^p}\lambda_L^q
+ M_2 K^{''}_{(pq)}(N_R^{pc})^TC^{-1}N_R^{qc}\bigr)
+{\rm h.c} \Bigr), \cr}\eqno(4.16)
$$
where we have denoted the family mixing matrices $K_{13}$, $K_{15}$
and $K_{56}$  by $K, K', K^{''} $, respectively. The symmetric and
antisymmetric parts of $K_{pq}$ are denoted by $K_{(pq)}$ and
$K_{[pq]}$, respectively.
Since we have three
neutral fields, $N_L$, $N_R^c $ and $\lambda_L$, and their mass eigenstates
are mixed, the mass matrix must be diagonalised. Ignoring the
mixing due to the generation matrices, the mass matix
of the neutral fields is of the form
$$
\bordermatrix{&N_L&N_R^c&\lambda_L \cr
N_L&0&m&0\cr N_R^c &m&M_2&M_3 \cr \lambda_L &0&M_3&0\cr},\eqno(4.17)
$$
and we shall assume a mass hierarchy $m\ll M_2,M_3$, and
$M_2\sim M_3$.
Diagonalisation of the matrix (4.17) produces two massive fields
whose masses are of order $M_2$, and the third will be a massless
left-handed neutrino.

The bosonic action can be read from Eq (3.9), for N=6.
The only complicated step is  the elimination
of the auxiliary fields, and one finds that the vev's used cannot
be arbitrary but must be related for the potential to survive
and the model to be consistent.
These relations are $M_G=M_1$ and $M_1M_2=-{K_{15}\overline{K_{15}}\over
2K_{12}K_{13}}M_3^2$. The bosonic action is
$$\eqalign{
&-4g^2 F_{\mu\nu}^{IJ}F^{\mu\nu IJ}+
2\vert K_{12}\vert^2 {\rm Tr}\Bigl(
\bigl( D_{\mu}({\m}+{\m}_0)\bigr)^2 \Bigr)
 +8\vert K_{13}\vert^2 {\rm Tr}\Bigl( \vert D_{\mu}({\n}+{\n}_0)
\vert^2 \Bigr) \cr
& +12\vert K_{15}\vert^2 \Bigl\vert D_{\mu}(H+H_0) \Bigr\vert^2
-V(\cal M,\cal N,H),\cr}\eqno(4.18)
$$
where the potential $V({\cal M}, {\cal N}, H)$ is
$$\eqalign{
& \bigl( {\rm Tr}\vert K_{12}
\vert^4-({\rm Tr}\vert K_{12}\vert^2)^2
\bigr) {\rm Tr}\Bigl( \vert {\m}+{\m}_0\vert^2 -\vert {\m}_0\vert^2
\Bigr)^2 \cr
&+4\Bigl\vert K_{13}K_{12}\bigl( ({\m}+{\m}_0)({\n}+{\n}_0)
+({\n}+{\n}_0)B(\overline {{\m}}+\overline {{\m}_0})B^{-1}
\bigr) \cr
&\qquad +2K_{15}\overline {K_{15}}\bigl( (H+H_0)B(\overline
H+\overline {H_0})\bigr) \Bigr\vert^2 \cr
& +8\Bigl\vert K_{12}K_{15} ({\m}+{\m}_0)(H+H_0)
+2K_{13}\overline {K_{15}}({\n}+{\n}_0)B(\overline H +\overline {H_0}
)\cr &\qquad -u(H+H_0)\Bigr\vert^2\cr
&+16\bigl( {\rm Tr}\vert K_{15}\vert^4 -({\rm Tr}\vert K_{15}\vert^2
)^2\Bigl\vert \vert H^* +H_0^*\vert^2 -M_3^2\Bigr\vert^2 \cr
&+16{\rm Tr}\vert K_{15}\vert^4 \Bigl\vert \vert H^* +H_0^*\vert^2
-M_3^2\Bigr\vert^2, \cr}\eqno(4.19)
$$
and $u=2K_{13}K_{15}\bigl( s+p-3(b+b')+2(a+a')+M_2\bigr)
-2K_{12}K_{25}M_1$.
We deduce that the SO(10) model is an attractive model. Its
construction is completely dictated by the arrangement of
the fermions, their representations, and the Dirac operator
acting on them. The nature of the Higgs fields is completely fixed,
and their vev's constrained by the requirement that the potential
is non-trivial for the consistency of the theory. The mass matrix
of the fermions is realistic.

{\bf\noindent 5. Gravity in non-commutative geometry }
\vskip.15truecm
\noindent
A natural question to ask is how to introduce gravity
in the framework of non-commutative geometry.
An answer to this question requires a generalisation
of the basic notions of
Riemannian geometry. Connes has proposed to define metric
proporties of a non-commutative space corresponding to an
involutive unital algebra $\cal A$ in terms of K-cycles over
$\cal A$ [2-3]. In [9] it was shown that every K-cycle over $\cal A$
yields a notion of "cotangent bundle" associated to $\cal A$ and
a Riemannian metric on the cotangent bundle. One can also introduce
analogues of the spin connection, torsion, Riemann curvature tensor,
Ricci tensor, and scalar curvature. This allows one to write the
generalized Einstein-Hilbert action. Here
we shall only describe the gravity action
for a two sheeted space, and refer the reader to [9] and
[4] for details. We shall also derive, heuristically,
an experimental signature of
the effect of the geometry on the standard model, which turns out
to be a constraint on the Higgs mass and top quark mass.

Consider a space-time $X$ consisting of two copies of a
four-dimensional manifold M: $X=M\times Z_2$. The algebra $\cal A$
is given by ${\cal A}=  C^{\infty}(M)\ot {\cal A}_1\bp C^{\infty}(M)\ot
{\cal A}_2$, where ${\cal A}_1={\cal A}_2 =C$.
The elements of ${\cal A}$ are operators of the form
${\rm diag}(1\ot a_1, 1\ot a_2)$ where $a_i$, $i=1,2 $ are
smooth function on $M$, and $1$ is the identity in the Clifford
algebra, ${\rm Cliff}(T^*M)$, of Dirac matrices over $M$.
We consider even K-cycles $(\pi ,H,D,\Gamma )$ for $\cal A$, with
$\pi =\pi_1 \bp \pi_2 $, where $\pi_i$ is a representation
of $C^{\infty }(M)\ot {\cal A}_i$ on a Hilbert space $L^2 (S_i,
\tau_i dv)$, where $S_i$ is a bundle of spinors on $M$ with
values in a finitely generated, projective hermitian left
${\cal A}_i $ module $E_i$, $\tau_i $ is a normalized trace on
${\cal A}_i$ and $dv$ is the volume element on $M$.
Then $h$ is defined by $h=L^2(S_1,\tau_1 ,dv)\bp L^2(S_2,\tau_2 ,dv)$.
The Dirac operator is taken to be
$$
D=\pmatrix{\Di_M \ot 1 &\g5\ot \phi \cr
\g5\ot \phi^* &\Di_M\ot 1\cr},\eqno(5.1)
$$
where $\Di_M$ is the standard covariant Dirac operator on $M$.
The $Z_2$ grading on $M$ is given by $\Gamma =\pmatrix{\g5 &0
\cr 0&-\g5 \cr}$. The "cotangent bundle" $\Omega_D^1 ({\cal A})
=Omega^1({\cal A})/{\rm ker}\pi $ is
a free left and right $\cal A$ module, with a basis $\{ e^N\}_{N=1}^{5}
$ given by
$$
e^a=\pmatrix{\gamma^a &0\cr 0&\gamma^a}, \qquad e^5=\pmatrix{
0&\g5\cr -\g5 &0}, \qquad a=1,2,3,4. \eqno(5.2)
$$
The hermitian structure on $\Omega_D^1 (\cal A)$ is given by the
trace of $8\times 8$ matrices, normalized such that ${\rm tr}1=1$.
Hence
$$
<e^N,e^M>={\rm tr}(e^N (e^M)^*)=\delta^{NM}. \eqno(5.3)
$$
For a one-form $\rho =\sum_i a_i db_i $ in $\Omega_D^1 (\cal A)$,
$\pi (\rho )$ is parametrized by
$$\pi (\rho )=\pmatrix{
\gamma^{\mu}\rho_{1\mu}&\g5\phi \rho_5\cr
-\g5\phi\tilde{\rho_5 }&\gamma^{\mu}\rho_{2\mu}\cr},\eqno(5.4)
$$
where $\rho_{1\mu}=\sum_i a_{i1}\partial_{\mu}b_{i1}$, $\rho_5
=\sum_i a_{i1}(b_{i2}-b_{i1})$, and similarly for $\rho_{2\mu}$ and
$\tilde{\rho_5}$. Evaluating $\pi (d\rho )=\sum_i [D,a_i][D,b_i]$,
we obtain
$$
\pi (d\rho )=-\pmatrix{g^{\mu\nu}\partial_{\mu}a_{i1}\partial_{\nu}
b_{i1}& 0\cr 0&g^{\mu\nu}\partial_{\mu}a_{i2}\partial_{\nu}b_{i2}\cr}.
\eqno(5.5)
$$
One sees that, for a suitable choice of $a_i, b_i$ subject to the
constraint $\pi (\rho )=0$, any expression of the form
${\rm diag }(X_1,X_2)$ can be obtained, where $X_1, X_2$ are
scalar functions. Therefore, we can express $\pi  (d\rho )$ modulo
auxiliary fields in terms of its components:
$$
\pi (d\rho )=\pmatrix{\gamma^{\mu\nu}\partial_{\mu}\rho_{1\nu}
& \phi \gamma^{\mu}\g5 (\partial_{\mu}\rho_5 +\rho_{1\mu}-\rho_{2\mu}\cr
-\phi \gamma^{\mu}\g5 (\partial_{\mu}\tilde{\rho_5}+\rho_{1\mu}-\rho_{2\mu})
&\gamma^{\mu\nu}\partial_{\mu}\alpha_{2\nu}\cr}.\eqno(5.6)
$$
This is a representative of
$\pi (d\rho )$ in $\pi \bigl( \Omega^2 ({\cal A})\bigr)
/ \pi \bigl( d{\rm Ker }\pi (\vert_{\Omega^1({\cal A})})\bigr) $
orthogonal to the auxiliary fields.
Let $\nabla $ be a connection on $\Omega_D^1 (\cal A)$
and $\omega^N_{\ M}\in \Omega_D^1 (\cal A)$ defined
by $\nabla e^N =-\omega^N_{\ M}\ot_{\cal A}e^M $. The components
of $\pi (\nabla )$ in the basis $\{ e^N\}_{N=1}^{N=5}$ are given
by
$$
\omega^N_{\ M}=\pmatrix{\gamma^{\mu}\omega^N_{1\mu M}&
\g5\phi l^N_{\ M}\cr -\g5\phi \tilde l^N_{\ M}&\gamma^{\mu}
\omega^N_{2\mu M}\cr}.\eqno(5.7)
$$
Hermiticity of $\nabla $ then implies that
$$
\omega^N_{i\mu M}=-\omega^M_{i\mu N}, \qquad i=1,2, \qquad
\tilde l^N_{\ M}=-l^M_{\ N}.\eqno(5.8)
$$
Let $T^N\in \Omega_D^2 (\cal A)$ be the components of the
torsion $T(\nabla )$ defined by $T^N =T(\nabla )e^N$. Then
$$
T^N =de^N +\omega^N_{\ M}e^M \eqno(5.9)
$$
Similarly define $R^N_{\ M}\in \Omega_D^2 (\cal A)$ by
$R(\nabla )e^N =R^N_{\ M}\ot_{\cal A} e^M$ where $R( \nabla )
$ is the Riemann curvature of $\nabla $ defined by $R(\nabla )
:=-\nabla^2 $. Then
$$
R^N_{\ M}=d\omega^N_{\ M}+\omega^N_{\ P}\omega^P_{\ M}.\eqno(5.10)
$$
Imposing the condition that the torsion $T(\nabla )$ vanishes
gives
$$\eqalign{
\omega^a_{1\mu b}&=\omega^a_{2\mu b}\equiv \omega^a_{\mu b},
\qquad \omega^a_{1\mu 5}=-\omega^a_{2\mu 5}=\phi l^a_{\ b}e^b_{\
\mu}, \cr
l^a_{\ b}&=l^b_{\ a}, \qquad l^5_{\ a}=-l^a_{\ 5}, \qquad
l^5_{\ a}e^a_{\ \mu}=-\partial_{\mu }\phi^{-1},\cr}\eqno(5.11)
$$
where $\omega^a_{\mu b}$ is the classical Levi-Civita connection
derived from the metric $g_{\mu \nu}=e_{\mu}^a\delta_{ab}e_{\nu}^b$
on $M$. The analogue of the Einstein-Hilbert action is
$$\eqalign{
I(\nabla )&:=\kappa^{-2}<R^N_{\ M}e^M,e_N> +\Lambda <1,1>\cr
 &=\kappa^{-2}\int_M {\rm tr} (R^N_{\ M}e^M (e_N)^*)
 +\Lambda \int 1,\cr}\eqno(5.12)
$$
where $\kappa^{-1}$ is the Planck scale.
This action is then calculated to be
$$\eqalign{
I(\nabla )&=\kappa^{-2}\int_M \Bigl( 2r-4\phi\nabla_{\mu}\partial^
{\mu}\phi^{-1}+4\phi^2 l^a_{\ a}l^5_{\ 5} \cr
&\qquad +\phi^2 \bigl( (l^a_{\ a})^2-l^a_{\ b}l^b_{\ a}\bigr)\Bigr)
\sqrt g d^4 x +2\Lambda \int_M \sqrt g d^4 x ,\cr}\eqno(5.13)
$$
where $r$ is the scalar curvature of the classical Levi-Civita
connection. The fields $l^a_{\ b}$ and $l^5_{\ 5}$ decouple, and
by setting $\phi =e^{-\kappa \sigma }$ one finds
$$
I(\nabla )=2\int_M (\kappa^{-2}r-2\partial_{\mu}\sigma \partial^{\mu}
\sigma +\Lambda )\sqrt g d^4 x . \eqno(5.14)
$$
Therefore a theory of gravity on $M\times Z_2$ is equivalent
to general relativity on $M$, with an additional massless
scalar field $\sigma $ that couples to the metric of $M$.
To better understand the role of the field $\sigma $ we can
study the coupling of gravity to the Yang-Mills sector [10]. In the
case of the standard model the field $\phi =e^{-\kappa \sigma }$
replaces the electroweak scale. In other words, the vev of the
field $\phi $ determines the electroweak scale. This simple
result has some unexpected consequences.
To determine the $\sigma $ dependence in the Yang-Mills action
of the standard model, we consider the $\sigma $ dependence in
the Dirac operator. For example, the leptonic Dirac operator
is
$$
D_l=\pmatrix{\gamma^a e_a^{\mu} (\partial_{\mu} +\ldots )\ot 1_2
\ot 1_3&\g5 \es \ot M_{12}\ot k\cr
\g5 \es \ot M_{12}^* \ot k^*&\gamma^a e_a^{\mu}(\partial_{\mu} +\ldots
)\ot 1_3\cr}.\eqno(5.15)
$$
{}From this one can easily verify that the bosonic part of the
standard model is
$$\eqalign{
L_b &= -{1\over 4} \Bigl ( F_{\mu\nu}^3 F^{\mu\nu 3}
+F_{\mu\nu}^2 F^{\mu\nu 2} +F_{\mu\nu}^1 F^{\mu\nu 1} \Bigl) \cr
&\qquad +D_{\mu} (H+M_{12})^* D_{\nu}(H+M_{12}) g^{\mu\nu}e^{-2\kappa
\sigma} \cr
&\qquad -{\lambda \over 24} \Bigl\vert \vert H+M_{12}\vert^2
-\vert M_{12}\vert^2 \Bigr\vert^2 e^{-4\kappa \sigma}. \cr}\eqno(5.16)
$$
The $\sigma $ dependence in Eq (5.16) is a consequence of the "Weyl
invariance" of the action (3.7) under  rescaling of
the Dirac operator
$D\ra e^{-w}D$, as this implies $g_{\mu\nu}\ra e^{2w}g_{\mu\nu}$ and
$\kappa\sigma \ra \kappa \sigma +w$. This can be easily seen from
the scalings: $\pi (\rho )\ra e^{-w}\pi (\rho )$ and
$\pi (\theta )\ra e^{-2w}\pi (\theta )$.
By redefining
$H+M_{12}\ra e^{\kappa\sigma} H$, the $H$ dependent terms in
(5.16) become
$$\eqalign{
D_{\mu}H^* D^{\mu}H +\kappa \partial_{\mu}(H^*H)\partial^{\mu}\sigma
+\kappa^2 H^*H \partial_{\mu}\partial^{\mu}\sigma
 -{\lambda \over 24}\Bigl\vert (H^*H)^2-\mu^2 e^{-2\kappa
\sigma}\Big\vert^2 .\cr}\eqno(5.17)
$$
The potential in Eq (5.17) could be rewritten in the familiar form
$$
V_0={\lambda \over 24} (H^*H)^2 -{1\over 2}m^2 (H^*H) +{3\over 2\lambda}
m^4, \eqno(5.18)
$$
where we have set $m^2 ={\lambda \mu^2\over 6}e^{-2\kappa\sigma}$,
so that $m$ is now a field and not just a parameter.
The potential $V_0$ is of the same form as that of the standard
model.
We  assume that,
after renormalization, the bosonic action takes the same form
as $I_l+I_q$.
In the absence of some understanding of
symmetries, it is not possible to prove this assumption at
the quanutm level.
Let $\phi $ be the component of the Higgs field that develops
a vev. We are then mainly interested in the potential
$$
V_0={\lambda \over 24} \phi^4 -{1\over 2}m^2\phi^2 +{3\over 2\lambda}
m^4 . \eqno(5.19)
$$
Minimizing with respect to $\phi $ and $m$ yields the
same asymmetric phase $\phi^2 ={6\over \lambda}m^2 $,
and the weak scale, $\es $, is undetermined at the classical level.
The quantum corrections to the potential are given, in the one-loop
approximation, by the
effective Coleman-Weinberg [11] potential of the standard model [12]:
$$\eqalign{
V_1&= {1\over 16\pi^2}\Bigl( {1\over 4}H^2 (\ln {H\over
M^2}-{3\over 2}) +{3\over 4}G^2 (\ln {G\over M^2} -{3\over 2})
+{3\over 2 }W^2 (\ln {W\over M^2}-{5\over 6})\cr
&\qquad \qquad +{3\over 4}Z^2 (\ln {Z\over M^2}-{5\over 6})
-3T^2 (\ln {T\over M^2}-{3\over 2})\Bigr), \cr}\eqno(5.20)
$$
where
$$\eqalign{
H&=-m^2 +{1\over 2} \lambda \phi^2 ,
\qquad G=-m^2 +{1\over 6} \lambda \phi^2 \cr
W&={1\over 4}g_2^2 \phi^2 ,\qquad Z={1\over 4}(g_2^2 +g_1^2)\phi^2 ,
\qquad T={1\over 2}h^2 \phi^2 ,\cr}\eqno(5.21)
$$
and $M$ is the renormalization scale.
Minimizing the total potential $V_0+V_1$ with respect to the
fields $\phi $ and $m$,  after rescaling
$$
G=\overline G M^2,\qquad H=\overline H M^2, \qquad T=\overline T M^2,
\eqno(5.20)
$$
the asymmetric solution  is given by the
solution to the following two equations:
$$
\eqalignno{
0 &=\overline G + {M^2\over 32\pi^2 \phi^2}(\overline H
-\overline G)\Bigl( \overline H (\ln \overline H -1)+3\overline G
(\ln \overline G -1)\Bigr), &(5.21) \cr
0 &=\overline G +{3M^2\over 32\pi^2\phi^2}(\overline H-\overline G)
\Bigl( \overline H (\ln \overline H -1)+\overline G (\ln \overline
G -1)\Bigr) -{g_2^2+g_1^2\over 64\pi^2}\cr
&\qquad +{3g_2^4\phi^2\over 128\pi^2M^2}\Bigl( \ln {g_2^2\phi^2
\over 4M^2}-{1\over 3}\Bigr) -{3M^2\over 4\pi^2\phi^2}
\overline T^2 (\ln \overline T-1). &(5.22) \cr}
$$
At the scale $M=m_Z$, the mass of the Z-particle, the coupling
constants $g_1,g_2$ as well as the vev $\phi $ are known
from experimental data, corrected with the help of the renormalization
group equations [12]:
$$
g_2=0.650, \qquad g_1=0.358 ,\qquad \phi =246 \ {\it Gev} .\eqno(5.23)
$$
The only unknowns in the minimization equations are $\lambda $,
$m$ and the square of the top quark mass $T=m_t^2$.
These equations, being complicated functions of $\overline H$ and
$\overline G$, can only be solved numerically, for various values
of $\overline T $. The numerical solutions are easily obtained using
{\it Mathematica}.
The Higgs mass can be determined
from the formula $m_H^2={\partial^2 V\over \partial \phi^2}$ which
gives
$$\eqalign{
m_H^2&=M^2\Bigl( (\overline H-\overline G)+{9M^2\over 16\pi^2
\phi^2}(\overline H-\overline G)^2(\ln \overline H +{1\over 3}
\ln \overline G )\cr
&\qquad \qquad +{3g_2^4\phi^2\over 64\pi^2 M^2}\ln {g_2^2\phi^2\over 4M^2}
\ -{3M^2\over 2\pi^2\phi^2}\overline T^2 \ln \overline T
\Bigr). \cr}\eqno(5.24)
$$
We now quote the results: There are only two classes of solutions,
for $\overline G\ll \overline H$ and for $\overline H\ll \overline G$.
In the first case, we find that there are only two narrow bands
for the top quark mass where  solutions exist. The first band is
$ 0.365\le \overline T \le 0.455, \quad \overline G\ll \overline H$,
corresponding to a top quark mass
$$
54.90\ {\rm Gev}\le m_t\le 61.35\ {\rm Gev},\eqno(5.25)
$$  which is already ruled out
experimentally. The second band is very narrow:
$2.57\le \overline T\le 2.61, \quad \overline G\ll \overline H$,
corresponding to the top quark mass
$$
146.23\ {\rm Gev}\le m_t\le 147.37 \ {\rm Gev},  \eqno(5.26)
$$
and a Higgs mass $117.26\ {\rm Gev}\le m_H\le 142.61\ {\rm Gev}$.
Clearly this band
of values for the top quark mass lies within the present experimental
average of [13]
$$
m_t=149 +\pmatrix{+21\cr -47\cr} \ {\rm Gev}. \eqno(5.27)
$$
The second class of solutions occurs when $1.30\le \overline T
\le 2.61, \quad \overline H\ll \overline G $,
corresponding to the top quark mass
$$
104.07\ {\rm Gev}\le m_t\le 147.48\ {\rm Gev}, \eqno(5.28)
$$
and a Higgs mass $1208\ {\rm Gev}\ge m_H\ge 1197\ {\rm Gev}$.
However, since
$\overline H\ll\overline G$, and since the coupling constant $\lambda
=O(-100)$, the potential, in this domain, becomes unbounded from below,
signaling the break down of the perturbative regime. Requiring
stability of the electroweak potential excludes this solution.
Therefore the only acceptable solution is (5.26) which is remarkably
constrained, considering the wide range of possibilities that one
might have, a priori.
We note that the field $\sigma $ becomes massive with the
square of the mass given by: $m_{\sigma}^2= {\partial^2 V\over \partial
\sigma^2} $. This is equal to
$$
m_{\sigma }^2 =\kappa^2 m^2 \Bigl( 2\phi^2 {\overline
H-4\overline G\over \overline H-\overline G} +{M^2\over 16\pi^2}
\bigl( \overline H (1-\ln \overline H)+3\overline G (1-\ln
\overline G)\bigr)\Bigr). \eqno(5.29)
$$
For the physically acceptable solutions we have $\overline H=O(1)$,
$\overline G=O(10^{-4})$ and $m^2=O(M^2)$. Then we find from Eq (5.29)
that
$$
m_{\sigma }^2=O(\kappa^2 M^4),  \eqno(5.30)
$$
so that $m_{\sigma} =O(10^{-15})$ Gev, which is unobservable.
These predictions have at best a heuristic value, since the
problem of fixing the form of the cosmological constant at the
one-loop level by imposing natural geometrical constraints is
not understood. However, they do suggest that gravitational
effects may play a role in understanding masses of fermions
and Higgses.

\vskip.5truecm
{\bf\noindent Acknowledgments}
\noindent
{\bf by A.H.C}: I would like to express my  indebtedness
and gratitude to
Abdus Salam for his role as my thesis advisor at Imperial
College, during the period 1974-1976, and as my mentor during 1977
when I stayed as a postdoctoral fellow in Trieste.  His quest
to understand the fundamental forces in nature, including
gravity, and his way of thinking have shaped my research since then.
I consider myself fortunate to have studied under such a
great physicist, teacher and a generous person, who found pleasure in
sharing his knowledge with the people around him.
\vskip1.0truecm
{\bf \noindent References}
\vskip.2truecm
\item{[1]} S. Glashow {\sl Nucl. Phys.} {\bf 22}, 579 (1961);\br
A. Salam and J. Ward {\sl Phys. Lett} {\bf 13}, 168 (1964);\br
S. Weinberg, {\sl Phys. Rev. Lett.} {\bf 19}, 1264 (1967);\br
A. Salam, in {\sl Elementary Particle Theory} (editor N. Svartholm),
Almquist and Forlag, Stockholm.

\item{[2]} A. Connes, {\sl Publ. Math. IHES} {\bf 62} 44 (1983);\br
 A. Connes, in {\sl the interface of mathematics
and particle physics }, Clarendon press, Oxford 1990, Eds
D. Quillen, G. Segal and  S. Tsou.

\item{[3]} A. Connes and J. Lott,{\sl Nucl.Phys.B Proc.Supp.}
{\bf 18B} 29 (1990), North-Holland, Amsterdam;
{\sl  Proceedings of
1991 Summer Cargese conference} p.53  editors J. Fr\"ohlich
et al (1992),Plenum Pub.;\br
D. Kastler , Marseille preprints
CPT-91/P.2610, CPT-91/P.2611 and P.2814;\br
For an alternative construction see \br
R. Coquereaux, G. Esposito-Far\'ese, G. Vaillant,
{\sl Nucl. Phys.} {\bf B353} 689 (1991);\br
M. Dubois-Violette, R. Kerner, J. Madore, {\sl J. Math.
Phys.} {\bf 31} (1990) 316;\br
B. Balakrishna, F. G\"ursey and K. C. Wali, {\sl Phys. Lett.}
{\bf 254B} (1991) 430;\br
H.-G. Ding, H.-Y. Guo, J.-M. Li and Ke Wu, Beijing preprint
ASITP-93-23.

\item{[4]}A. Connes, {\sl Non-Commutative Geometry}, Academic press,
to be published;\br
D.Kastler and M. Mebkhout, {\sl Lectures on non-commutative
geometry and applications to elementary particle physics},
to be published;\br
A.H. Chamseddine and J. Fr\"ohlich, {\sl Some elements of Connes'
non-commutative geometry and space-time physics}, Zurich preprint
ZU-TH-17/1993

\item{[5]} A.H. Chamseddine, G. Felder and J. Fr\"ohlich,
{\sl Phys. Lett.} {\bf296B} (1993) 109; {\sl Nucl. Phys.}
{\bf B395} (1993) 672.

\item{[6]} E. Alvarez, J.M. Garcia-Bondia and C.P. Martin,
{\sl Madrid preprint}, May 1993.

\item{[7]} H. Georgi in {\sl Particles and Fields} {\bf 1974} p.575,
editor C.E. Carlson (AIP, New York, 1975);\br
H. Fritsch and P. Minkowski, {\sl Ann. Phys.} (N.Y) {\bf
93}  (1977) 193;\br
M.S. Chanowitz, J. Ellis and M. Gaillard, {\sl Nucl. Phys.} {\bf
B129} (1977) 506.

\item{[8]}A.H. Chamseddine and J. Fr\"ohlich, {\sl SO(10)
unification in non-commutative geometry},
Zurich preprint ZU-TH-10/1993.

\item{[9]} A.H. Chamseddine, G. Felder and J. Fr\"ohlich,
{\sl Comm.Math.Phys} {\bf 155} (1993) 205.

\item{[10]}A.H. Chamseddine and J. Fr\"ohlich, {\sl Constraints
on the Higgs and top quark masses from effective potential and
non-commutative geometry}, {\sl Phys. Lett.} {\bf B314} (1993) 308.

\item{[11]} E. Weinberg and S. Coleman, {\sl Phys.Rev.} {\bf D7}
(1973) 1888.

\item{[12]}M. Sher, {\sl Phys.Rep.} {\bf 179} (1989) 274 and
references therein;\br
C. Ford, D.R.T. Jones, Y.W. Stephenson and M.B. Einhorn,
{\sl Liverpool preprint}  LTH 288.

\item{[13]} U. Amaldi, This Proceedings.

\end